\documentclass[aps,prb, reprint, superscriptaddress]{revtex4-2}

\usepackage{graphicx}
\usepackage{dcolumn}
\usepackage{bm}
\usepackage{nicefrac}
\usepackage{tabularx}

\usepackage{floatrow}
\usepackage{siunitx}

\usepackage{amsmath}

\usepackage{amsmath,amssymb,bm}
\usepackage{graphicx}
\usepackage{xspace}
\usepackage{latexsym}
\usepackage{color}
\usepackage{hyperref}
\usepackage{placeins}

\usepackage[normalem]{ulem}

\usepackage{booktabs}
\usepackage{multirow}
\usepackage{tabularx}
\usepackage{nicematrix}

\usepackage{dsfont}
\usepackage{braket}
\usepackage{bbold}

\newcommand{\SrZn}{SrZnVO(PO$_4$)$_2$\xspace}

\newcommand{\BaCd}{BaCdVO(PO$_4$)$_2$\xspace}

\newcommand{\PbV}{Pb$_2$VO(PO$_4$)$_2$\xspace}

\newcommand{\be}{\begin{equation} }
	\newcommand{\ee}{\end{equation} }
\newcommand{\bea}{\begin{eqnarray} }
	\newcommand{\eea}{\end{eqnarray} }

\newcommand{\mb}[1]{\mathbf{#1}}

\newcommand{\meV}{\milli\electronvolt}
\newcommand{\VV}{V$^{4+}$\xspace}

\newcolumntype{Y}{>{\centering\arraybackslash}X}

\newcommand{\CAFa}{CAF$_\text{a}$\xspace}
\newcommand{\CAFb}{CAF$_\text{b}$\xspace}

\begin{document}
	
	
	\title{Spin correlations in the frustrated ferro-antiferromagnet \SrZn near saturation.}
	
	
	\author{F. Landolt}
	\email[]{landoltf@phys.ethz.ch}
	\author{K. Povarov}
	\author{Z. Yan}
	\author{S. Gvasaliya}
	\affiliation{Laboratory for Solid State Physics, ETH Zürich, 8093 Zürich, Switzerland}
	
	\author{E. Ressouche}
	\author{S. Raymond}
	\affiliation{Université Grenoble Alpes, CEA, IRIG, MEM, MDN, 38000 Grenoble, France}
	
	\author{V. O. Garlea}
	\affiliation{Neutron Scattering Division, Oak Ridge National Laboratory, Oak Ridge, Tennessee 37831, USA}

	\author{A. Zheludev}
	\homepage{http://www.neutron.ethz.ch/}
	\affiliation{Laboratory for Solid State Physics, ETH Zürich, 8093 Zürich, Switzerland}

	
	\date{\today}
	
	\begin{abstract}
		Single crystal elastic and inelastic neutron scattering experiments are performed on the frustrated ferro-antiferromagnet \SrZn in high magnetic fields.
		The fully polarized state, the presaturation phase and the columnar-antiferromagnetic phase just bellow the presaturation phase were investigated. The observed renormalization of spin wave bandwidths, re-distribution of intensities between different branches and non-linearities in the magnetization curve are all indicative of strong deviations from classical spin wave theory. The previously observed presaturation transition is attributed to a staggered pattern of Dzyaloshinskii-Moriya interactions.
	\end{abstract}
	
	
	\maketitle
	
	\section{Introduction} \label{sec:Intro}
	
	The layered vanadyl phosphates ABVO(PO$_4$)$_2$ (A,B = Sr, Zn, Pb, Ba, Cd) are hailed as proximate realizations of the conceptually important $J_1-J_2$ ferro-antiferromagnetic square lattice model \cite{Nath2008, Tsirlin2009, TsirlinRosner2009, Bossoni2011}. The main interest stems from the spin-nematic state  that was predicted to emerge in this model in applied magnetic fields, at some transition field $H_c$ just below the saturation field $H_\text{sat}$  \cite{Shannon2006, Shindou2009, Smerald2015}. Indeed, all three most studied compounds of the series, namely \BaCd \cite{Povarov2019,Bhartiya2019, Skoulatos2019,Bhartiya2021}, \SrZn \cite{Landolt2021} and \PbV\cite{Bettler2019,Landolt2020}, show unusual presaturation behavior. In the two latter materials the existence of a well-defined presaturation phase is confirmed beyond any doubt.  However, recent NMR studies  \cite{ranjith2021nmr,Landolt2020} clearly show that it is not a spin nematic phase (quadrupolar order) but has spontaneous time-reversal symmetry breaking. Moreover, coming from the fully saturated (paramagnetic) state, the presaturation phase in \SrZn \cite{ranjith2021nmr} emerges in a single-magnon condensation process. This contrasts with the condensation of 2-magnon bound states needed to produce a spin-nematic phase. To date the nature and origin of this presaturation phase remain unresolved.
	
	In the present work we report a series of neutron diffraction and inelastic neutron scattering studies in high magnetic fields to address this lingering mystery. Our results indicate that the transition at $H_c$ involves a peculiar type of spin-reorientation that is of purely classical origin and is caused by Dzyaloshinskii-Moriya interactions. At the same time we find that quantum fluctuations are relevant in this system. They strongly influence the magnetization process and lead to a huge renormalization of spin wave bandwidths.

	\begin{figure*}
		\includegraphics[width=\columnwidth]{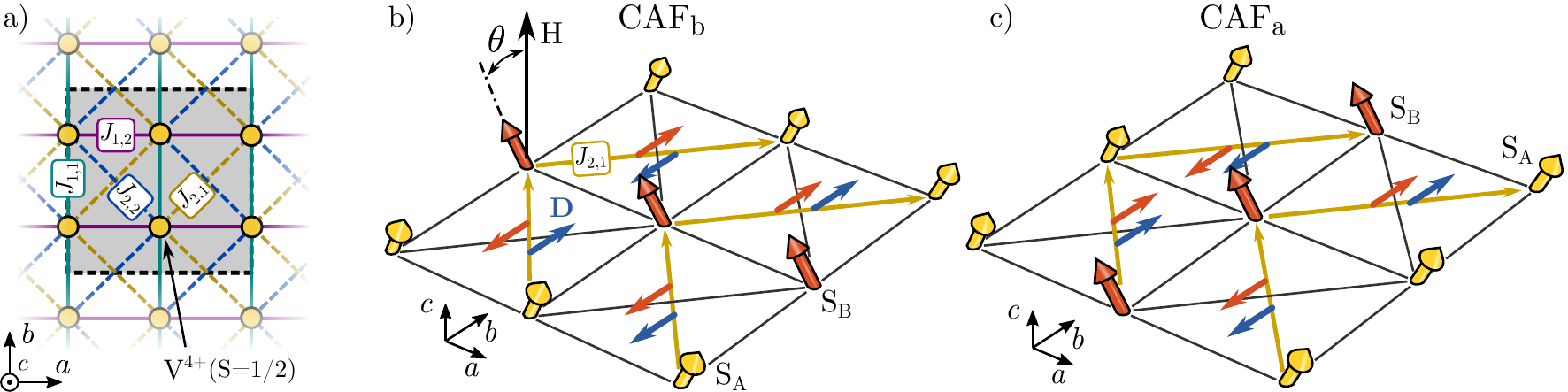}
		\caption{a) Heisenberg exchange constants in the \VV layers of \SrZn. Yellow circles are \VV ions. The grey rectangle represents the unit cell. b) A partially magnetized \CAFb~type spin structure with canting angle $\theta$ and the AF moment pointing along $\mathbf{a}$. The blue arrows indicate a crystal symmetry-compatible Dzyaloshinskii vector on the $J_{2,1}$ bond. Orange arrows shows the cross products of the respective interacting spins. c) The same as in b) but for the \CAFa~spin structure.}
		\label{fig:DM_Frustration}
	\end{figure*}

	The crystal structure and Heisenberg spin Hamiltonian for \SrZn are discussed in detail in \cite{SrZn_Struct,Landolt2021}. Only the key points are summarized here. The material is orthorhombic ($Pbca$) with lattice parameters $a = 9.066(1)$~\AA, $b=9.012(1)$~\AA~and $c = 17.513(1)$\AA.  The $S=1/2$ \VV ions are arranged in layers parallel to the $(a,b)$ plane with negligibly weak magnetic interactions along the $c$ direction. Within each unit cell, each layer has four magnetic \VV ions. The crystal symmetries allow for two nearest-neighbor (nn) $J_{1,1}$ and $J_{1,2}$ and two next-nearest-neighbor (nnn) $J_{2,1}$ and $J_{2,2}$ exchange constants. The coupling geometry is shown in Fig.~\ref{fig:DM_Frustration}~(a). The corresponding Heisenberg model was shown to reproduce the spin wave spectrum in zero field rather well, with all nn couplings negative (ferromagnetic, FM) and all nnn ones positive (antiferromagnetic, AFM). The previously reported exchange parameters obtained from fits to the measured neutron spectra are summarized in the 2nd column in Table~\ref{tab:J}. In zero applied field the material orders magnetically at $T_\text{N} = \SI{2.6}{\kelvin}$ in a so-called columnar antiferromagnetic (CAF) structure.  The sublattice spins are primarily along the $\mathbf{a}$ axis, aligned anti-parallel to their nearest neighbors along the $\mathbf{b}$  axis and parallel to those along  $\mathbf{a}$ \cite{Landolt2021}. Below we shall refer to this structure as ``\CAFb''. It is fully consistent with the exchange constants, the $\mathbf{a}$-axis ferromagnetic coupling $J_{1,2}$ being stronger than $J_{1,1}$ along the $\mathbf{b}$ axis. At low temperature, in magnetic fields applied along the $c$ axis, the \CAFb phase survives up to a discontinuous transition at $\mu_0 H_c=13.63$~T\cite{Landolt2021}. The presaturation phase (PS) that follows extends up to full saturation (SAT) at $\mu_0 H_\text{sat}=14.06$~T~\cite{ranjith2021nmr}.
	
	\section{Experimental} \label{sec:Exp}
	All experiments reported below were performed on \SrZn single crystal samples grown by the same process as those used in \cite{Landolt2021}. In all cases the external magnetic field was applied along the crystallographic $\mathbf{c}$ axis. For reference, the corresponding gyromagnetic ratio for this geometry is $g=1.926$ \cite{Forster2013}.

	\paragraph{Magnetization} of \SrZn was measured at $T=150$~mK on a \SI{0.2}{\milli\gram} single crystal sample using a custom-built Faraday force magnetometer \cite{Blosser2020} in fields of up to 14~T.
	
	\begin{figure}
		\includegraphics[width=\columnwidth]{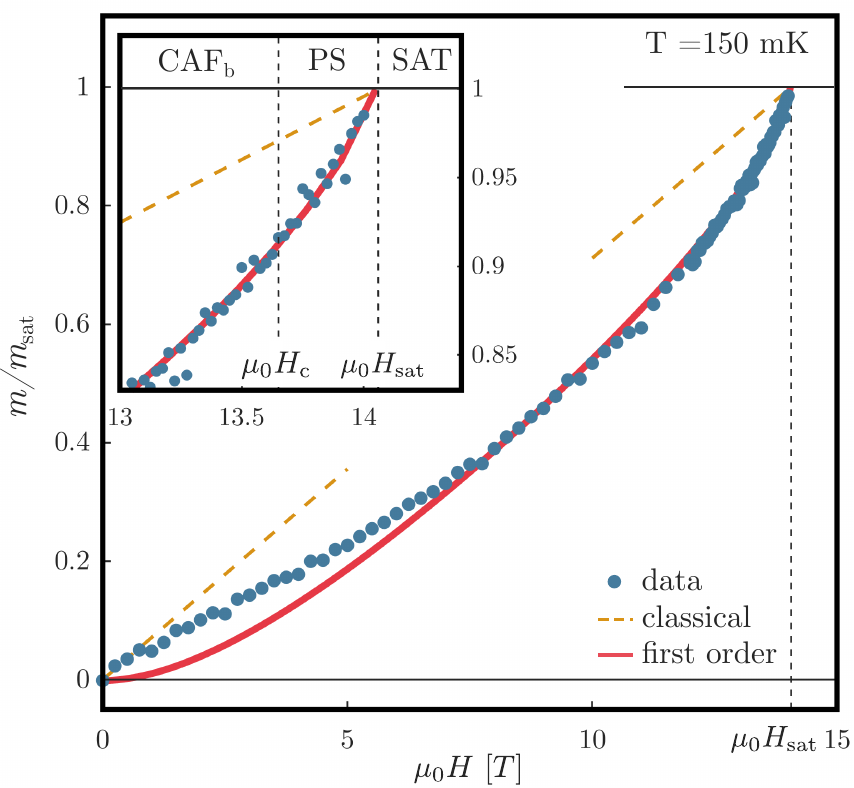}
		\caption{Measured magnetization in a \SrZn single crystal in magnetic fields applied along $\mathbf{c}$ (symbols). The orange dashed line is the linear-SWT result. The red solid line includes 1st-order $1/S$ corrections for the $J_1-J2$ model following \cite{Thalmeier2008}.}
		\label{fig:Magnetization}
	\end{figure}

	\paragraph{Neutron diffraction} was carried out on the D23 lifting counter diffractometer at ILL using a pyrolytic graphite (PG) monochromator to produce a  $E_i=\SI{14.7}{\milli\electronvolt}$ incident beam. Sample environment was a 15~T cryomagnet and $^3$He-$^4$He dilution refrigerator. All data were taken at temperatures of about 75~mK. An 80~mg \SrZn single crystal sample was mounted with the $\mathbf{c}$-axis vertical. Measurements of magnetic Bragg intensities in \SrZn close to saturation are extremely challenging. On the one hand, the ordered moment becomes very small in this field range, while Bragg intensities scale as its square. On the other hand, a $\mathbf{Q}=0$ propagation vector of the \CAFb structure implies that most magnetic Bragg peaks are located on top of much stronger nuclear reflections. The only exceptions in the scattering plane are forbidden nuclear reflections of type $(h,k,0)$ with $h$-odd and $(0,k,0)$ with $k$-odd. Even for these reciprocal-space points the background is rather high due to double-scattering and/or higher order beam contamination. In the present study we only followed the peak intensities of a handful of magnetic reflections as a function of applied field. The typical cumulative counting time was about 30~min/point.
	
	\paragraph{Neutron spectroscopy} experiments were performed on a $\approx600$~mg sample with a mosaic spread of about $1^\circ$. Data in the ordered state just outside (at $\mu_0H=13.4~T$) and just inside (at $\mu_0H=13.8~T$) the presaturation phase were measured using the time of flight (TOF) spectrometer HYSPEC at SNS, ORNL with a 14~T cryomagnet and a dilution refrigerator. The incident neutron beam at HYSPEC is monochromated using a Fermi chopper and is then focused onto the sample using a PG monochromator. The Fermi choppers were operated at \SI{360}{\hertz} and was set to $E_i=\SI{5.3}{\milli\electronvolt}$ resulting in an energy resolution (FWHM) of \SI{0.16}{\milli\electronvolt}. Scattering events were recorded while rotating the sample over 145° in a step of 0.5°, with a counting time of \SI{8}{\minute} per step.
	
	Neutron spectra in the fully saturated state were measured on the same \SI{0.6}{\gram} sample using the triple axis spectrometer IN12 at ILL. A a dilution refrigerator and a 15~T cryomagnet were employed. The instrument was operated in a fixed final energy mode with $E_f = \SI{3.7}{\milli\electronvolt}$, with double-focusing PG monochromator and PG analyzer. Higher order beam contamination was suppressed by a velocity selector. An 80' collimator was mounted after the monochromator. The resulting energy width of the elastic line was measured to be \SI{0.22}{\milli\electronvolt} at FWHM. Data were collected in energy scans at fixed momentum transfers, while counting for about \SI{250}{\second} per point.

	%

	\begin{figure}
		\includegraphics[width=\columnwidth]{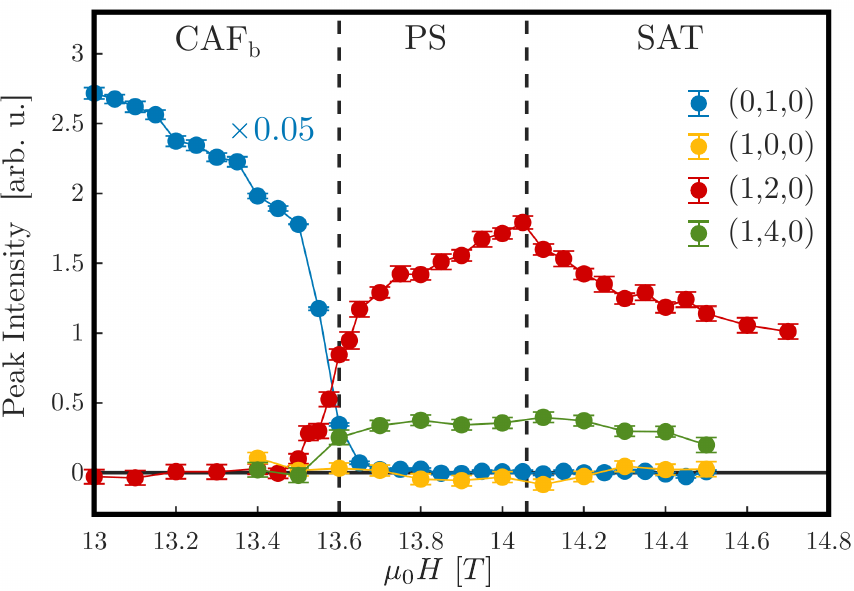}
		\caption{ Peak intensities of several potentially magnetic neutron Bragg peaks  measured as a function of magnetic field applied along the $\mathbf{c}$ axis. The background due to multiple scattering and higher order beam contamination has been subtracted as described in the text. }
		\label{fig:D23_FieldScan}
	\end{figure}

	\begin{figure*}
		\includegraphics[width=\columnwidth]{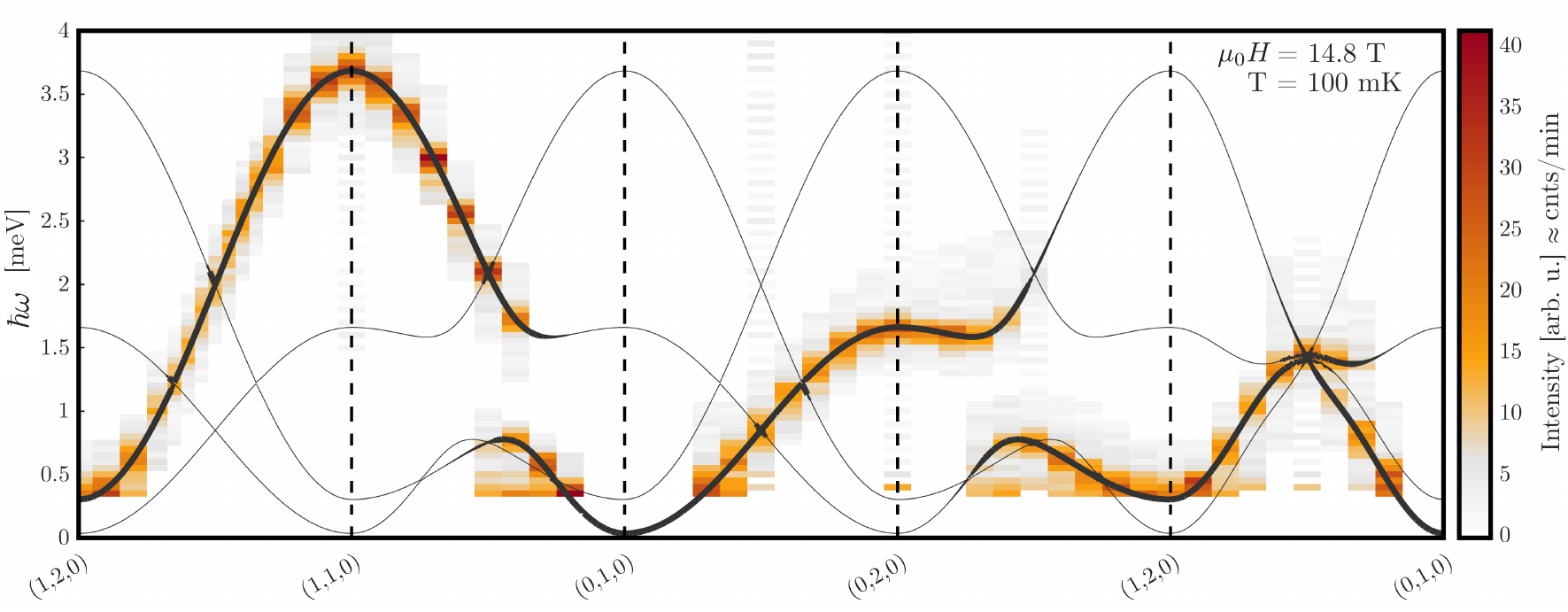}
		\caption{\label{fig:IN12_map} False color plot of neutron scattering intensities measured in \SrZn just above saturation. Solid line is a linear SWT fit as described in the text.}
	\end{figure*}
	
	\begin{figure}
		\includegraphics[width=\columnwidth]{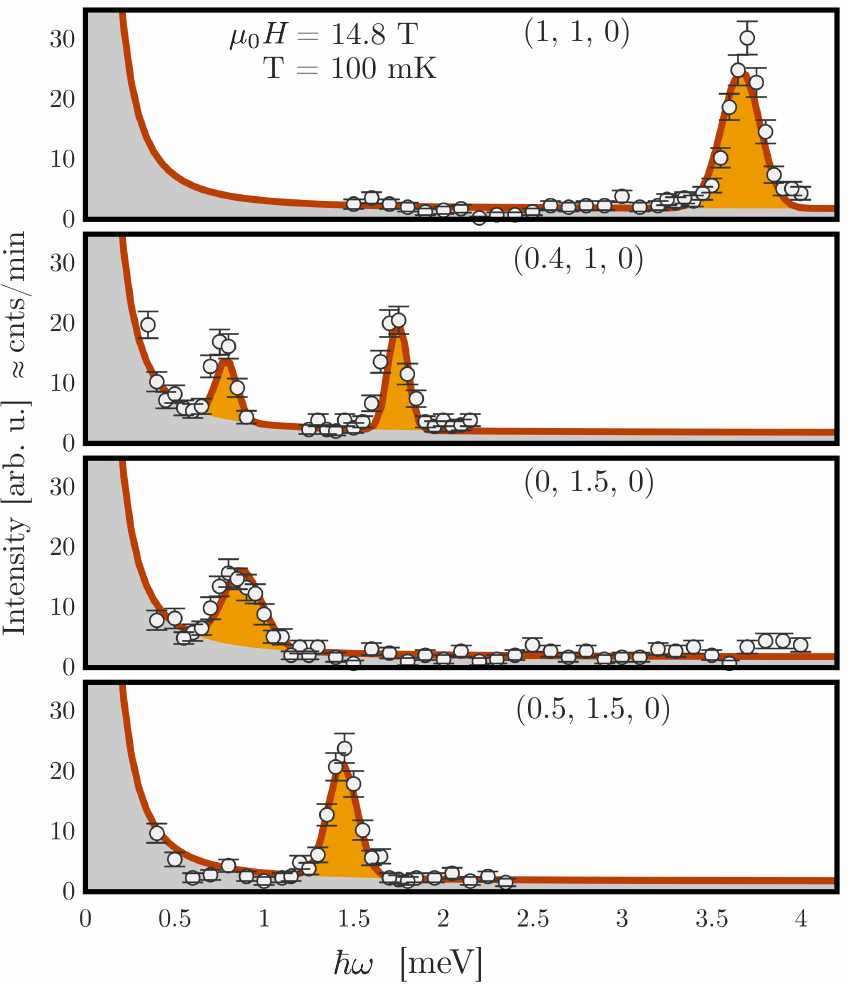}
		\caption{\label{fig:IN12_scans} Typical constant-$q$ scans  measured in \SrZn at $\mu_0H=14.8$~T and $T=100$~mK, just above the saturation transition (symbols). The solid lines are results of a global fit to the data as described in the text. Gray area is the modeled background contribution.}
	\end{figure}

	\section{Results} \label{sec::ExpRes}

	\subsection{Magnetization}
	The magnetization curve measured at \SI{150}{\milli\kelvin} is shown in Fig. \ref{fig:Magnetization} (symbols). The data shows a strong nonlinear behavior with a distinctive but non-divergent upturn toward saturation. No visible footprint of the transition at $H_c$ can be recognized in the data.

	\subsection{Magnetic Bragg scattering}
	The peak intensities of four purely magnetic Bragg reflections  $(0,1,0)$, $(1,0,0)$, $(1,2,0)$ and $(1,4,0)$, respectively, measured as a function of field at $T\approx75$~mK  are shown in Fig. \ref{fig:D23_FieldScan}. Here the background is subtracted. For $(0,1,0)$, a characteristic reflection of the zero-field \CAFb phase \cite{Landolt2021}, the background was measured in the saturated phase. For the other three peaks that are absent in the \CAFb phase, it was taken below $H_c$. Two important features are to be recognized. i) At just about $H_c$ the magnetic scattering intensity disappears abruptly  at $(0,1,0)$, as observed previously. At the same time, a magnetic contribution appears at the $(1,2,0)$ and $(1,4,0)$ positions. Note, however, that the intensity of those reflections is an order of magnitude smaller than that lost in $(0,1,0)$. Note also that the $(1,0,0)$ peak does not appear. ii) The intensity at the $(1,2,0)$ position reaches a maximum at saturation field $H_\text{sat}$. Beyond that point the intensity decreases progressively. $(1,4,0)$ does not show any obvious feature at $H_\text{sat}$.

	\subsection{Magnetic excitations: fully polarized state}
	A summary of the inelastic neutron scattering data measured in the fully polarized state at \SI{14.9}{\tesla} is shown in Fig.~\ref{fig:IN12_map} as a false color intensity plot. The data were obtained in constant-$Q$ scans, such as those shown in Fig.~\ref{fig:IN12_scans}. The data were analyzed very similarly to the way it was done in  \cite{Landolt2021} for zero field. Intensities were calculated within linear spin wave theory (SWT) based on a Heisenberg Hamiltonian with four exchange constants. The latter were additionally constraint to give the correct value of the saturation field $g \mu_B \mu_0 H_{\text{sat}} = 4S(J_{1,1}+J_{2,1}+J_{2,2})$. The neutron intensities were  calculated using the library spinW \cite{Toth2015} and folded numerically with experimental resolution computed in the Popovici approximation using the ResLib package \cite{Reslib}. The magnetic form factor for \VV was taken from \cite{ITC_VC}. At all wave vectors the background was a Gaussian centered at zero energy transfer to represent elastic-incoherent and quasielastic scattering plus a flat (energy-independent) contribution. Fitting this 7-parameter  model (3 exchange constants, an overall scale factor, elastic line height and width and background) {\em globally} to all data collected achieved a weighted squared error of $\chi^2=2.4$. The fitted values for the exchange parameter are tabulated in Tab.~\ref{tab:J} for a direct comparison with values obtained in zero field \cite{Landolt2021}. The calculated spin wave dispersion is plotted as black lines in Fig.~\ref{fig:IN12_map} with line thickness proportional to scattering intensity. Scans simulated based on the obtained fit parameters are compared to the data in  Fig.~\ref{fig:IN12_scans}.

	\begin{table}
			\def\arraystretch{1.4}
			\newcolumntype{s}{>{\hsize=.3\hsize \centering}X}
			\newcolumntype{Y}{>{\centering\arraybackslash}X}
			\newcolumntype{b}{>{\hsize=.1\hsize \centering}X}
			\begin{tabularx}{1.0\textwidth}{ s | Y Y c}
				\hline
				\hline
				& \SI{14.9}{\tesla} & \SI{0}{\tesla} & $J_{\text{0T}} / J_{\text{14.9T}}$\\
				\hline
				$J_{1,1}$ & $-\SI{0.45(1)}{\meV}$ & $-\SI{0.35(1)}{\meV}$ &  0.78 \\
				$J_{1,2}$ & $-\SI{0.56(1)}{\meV}$ & $-\SI{0.42(1)}{\meV}$ &  0.75 \\
				$J_{2,1}$ & $+\SI{0.86(1)}{\meV}$ & $+\SI{1.21(1)}{\meV}$ &  1.41 \\
				$J_{2,2}$ & $+\SI{0.37(1)}{\meV}$ & $+\SI{0.32(1)}{\meV}$ &  0.86 \\
				\hline
				\hline
			\end{tabularx}
			
			\caption{Heisenberg exchange parameters obtained by fitting a linear spin wave model to inelastic neutron data measured in the fully saturated phase at \SI{14.9}{\tesla} and in zero field\cite{Landolt2021}. The last column displays the renormalization of the zero field exchange parameters compared to their values obtained in the saturated phase.}
			\label{tab:J}
		\end{table}

		\subsection{Magnetic excitations: below saturation}
\begin{figure*}[]
			\includegraphics[width=\columnwidth]{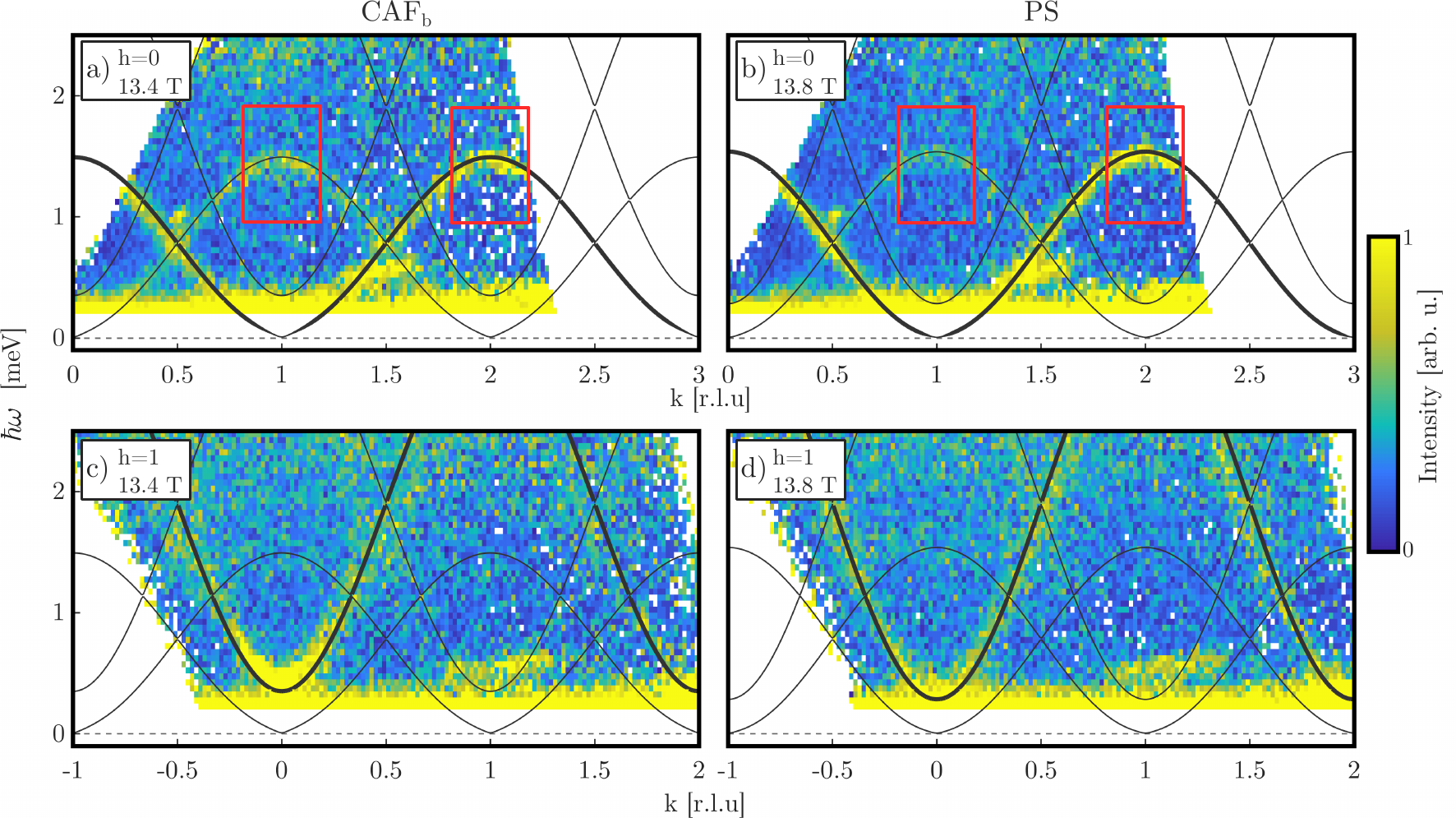}
			\caption{\label{fig:HYSPEC_map} Neutron scattering intensities measured in the TOF experiment. The false color plots are energy-momentum slices along $(0,k,0)$ (a,b) and $(1,k,0)$ (c,d). The data is fully integrated along $l$ and $\pm 0.2$~r.l.u. along the $h$-direction. The data are taken just below (a,c) and just above (b,d) the presaturation transition at $H_c$, at $T=250$~mK.  The red rectangles indicate the data used for the constant-Q cuts shown in \ref{fig:HYSPEC_QCuts}.}
		\end{figure*}

\begin{figure}[]
			\includegraphics[width=\columnwidth]{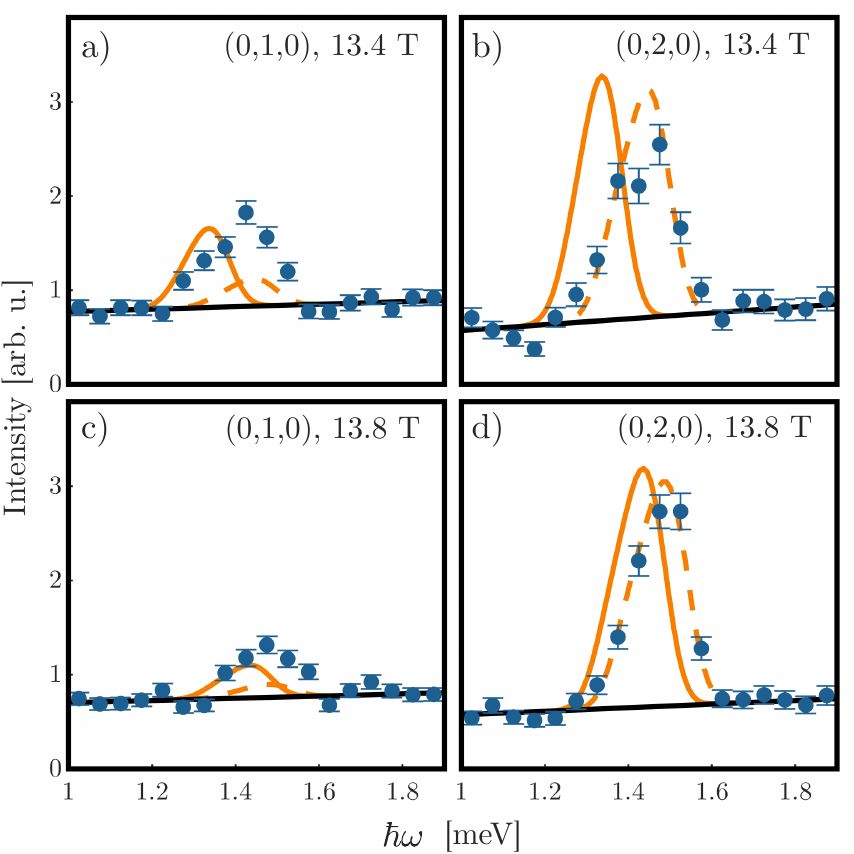}
			\caption{Symbols: energy scans cut from the TOF data obtained by momentum-integration of intensities in areas bordered in red in Fig.~\ref{fig:HYSPEC_map}. The dashed line is a SWT simulation based on exchange constants measured above saturation. The solid colored line is the same simulation performed at a fictitious value of magnetic field in order to obtain the correct magnetization value in SWT. The solid straight line is a linear background estimate.}
			\label{fig:HYSPEC_QCuts}
		\end{figure}

		Typical energy-momentum slices through the TOF data measured at \SI{13.4}{\tesla} and \SI{13.8}{\tesla} are shown in Fig.~\ref{fig:HYSPEC_map} as false color intensity plots.
		They correspond to momentum transfers along the $(0,k,0)$ and $(1,k,0)$ reciprocal-space lines. In all cases the intensity is integrated fully along $l$ and in a range $\pm 0.2$~[r.l.u.] along $h$. 	
		To within experimental resolution and in the energy range not affected by the strong elastic incoherent scattering, there seem to be no qualitative differences between spectra collected below and above $H_c$, respectively.
		Unlike above saturation, two excitation branches are distinctly visible.  The weaker of the two appears to further lose weight as the saturation field is approached, and is no longer visible beyond saturation (see Fig.~\ref{fig:IN12_map}).
		For a better comparison of the intensities in the two branches, in Fig~\ref{fig:HYSPEC_QCuts} we show energy-cuts obtained by additionally integrating the data in the range  $k=1\pm0.2$ and $k=2\pm0.2$, as indicated by the red rectangles in Fig.~\ref{fig:HYSPEC_map}. Assuming the linear background shown in  Fig~\ref{fig:HYSPEC_QCuts}, the respective integrated intensities of both modes are tabulated in Tab.~\ref{tab:Intensities}.
		
		The dispersions of both spin wave branches are well reproduced by the SWT model and exchange constants determined in the polarized state. These calculations are shown in  Fig.~\ref{fig:HYSPEC_map}  in solid lines. Discrepancies are revealed only in a closer look at the intensities. Consider the dashed lines in Fig.~\ref{fig:HYSPEC_QCuts}. They represent an SWT calculation of the two modes with an intensity scale factor to match the observed integrated intensity at $(0,2,0)$. The peak simulated  with the same scale factor   at $(0,1,0)$ appears considerably weaker than the observed scattering. We conclude that SWT fails to correctly reproduce the intensity distribution between the two modes even very close to full saturation.
		
		During the experiment also a large amount of elastic data was collected. They cover several complete Brillouin zones but show {\em no new magnetic reflections with propagation vectors other than $\mathbf{Q}=0$ appearing in the presaturation state}. Any redistribution of intensity among integer-index ($\mathbf{Q}=0$) Bragg position as seen on D23 is not possible to analyze due to considerable multiple scattering in those positions.

		\begin{table}
			\def\arraystretch{1.4}
			\newcolumntype{s}{>{\centering\arraybackslash\hsize=1.0\hsize}X}
			\newcolumntype{Y}{>{\centering\arraybackslash}X}
			\newcolumntype{Z}{>{\centering\arraybackslash\hsize=1.0\hsize}X}
			
			\begin{tabularx}{\textwidth}{Y Y  Y Y Y}
				\hline
				\hline
				position& field & observed & calc. (correct~field) & calc. (correct~magn.)\\[-0.2em]
				\hline
				\multirow{2}{*}{(0,2,0)} & 13.8~T & 1.00(6) & 1.00 & 1.00 \\
				& 13.4~T & 0.92(8) & 0.97 & 0.92 \\
				\hline
				\multirow{2}{*}{(0,1,0)} & 13.8~T & 0.27(4) & 0.06 & 0.14 \\
				& 13.4~T & 0.46(5) & 0.11 & 0.30\\
				\hline
				\hline
			\end{tabularx}
			\caption{Comparison between observed and calculated energy-integrated intensities of the inelastic peaks shown in Fig.~\ref{fig:HYSPEC_QCuts}, as discussed in the text. The calculated values have been normalized to match the observed intensities at $(0,2,0)$ measured at $13.8$~T.}
			\label{tab:Intensities}
		\end{table}
		
		\section{Discussion}
		
		\subsection{The presaturation transition}

		As mentioned, NMR experiments indicate that coming from the fully saturated phase, $H_\text{sat}$ exactly corresponds to the closure of a single-magnon gap \cite{ranjith2021nmr}. Furthermore, our experiments show that in the fully polarized state the observed intensities (magnon structure factors) are fully consistent with SWT. In the latter, the lowest-energy magnon in the saturated state (where SWT is exact) always corresponds to a spin correlation pattern that minimizes the classical exchange energy. That is, to the \CAFb state in the case of \SrZn. It is this magnon that can be expected to condense at $H_\text{sat}$ in a ``magnon-BEC'' transition \cite{Batyev1984}. If we were indeed dealing with a quantum Heisenberg spin Hamiltonian, we would necessarily recover  the \CAFb phase just below $H_\text{sat}$. Experimentally this is not the case:   $([\text{even}],[\text{odd}],[\text{even}])$ reflections characteristic of \CAFb are absent or at least much suppressed there. The logical conclusion is that the Heisenberg model is not the whole story and that the presaturation phase is stabilized by some additional terms in the Hamiltonian, such as anisotropy. They would have to be rather small, of the order of a few percent of the exchange constants, to remain undetected within the energy resolution of our inelastic experiments.
		
		The discontinuous collapse of the \CAFb structure at $H_c$ is accompanied by the appearance of new  $([\text{odd}],[\text{even}],[\text{even}])$ reflections. That would indicate that nearest neighbor spins along the $b$ axis become aligned parallel to one another, while those along $a$ are anti-parallel. In our notation, this is the \CAFa structure shown in Fig.~\ref{fig:DM_Frustration}~(c), and clearly contradicts the established hierarchy of exchange coupling constants, where $|J_{1,2}|>|J_{1,1}|$. The systematic absence of the $(1,0,0)$ reflection indicates that the moments of the \CAFa correlations point along $\mathbf{a}$, which suppresses the $(1,0,0)$ reflection due to the neutron polarization factor. It is crucial to emphasize that the newly appearing reflections {\em do not represent the magnetic order parameter of the presaturation phase}. Indeed they {\em persist in the paramagnetic state} and thus don't represent any spontaneous symmetry breaking. Since there is a true thermodynamic transition at $H_\text{sat}$, and also since the rather large intensity of $(0,1,0)$ below $H_c$ is not recovered above, we conclude that magnetic Bragg peaks corresponding to the order parameter of the presaturation phase must be present elsewhere in reciprocal space, but have eluded detection.
		
		While we lack the information regarding the order parameter of the presaturation phase, we can conceive of a toy model that illustrates  how even a {\em weak} Dzyaloshinskii-Moriya anisotropy can lead to a ``premature'' collapse of the \CAFb phase in favor of a \CAFa-type spin arrangement.
		Lets assume (somewhat arbitrarily) that there is a Dzyaloshinskii vector $\mathbf{D}\|\mathbf{b}$ associated with each $J_{2,1}$ bond. According to the symmetry analysis of Ref.~\cite{Landolt2021} such a Dzyaloshinskii vector will be sign-alternating between subsequent $J_{2,1}$ bonds along both the $\mathbf{a}$ and $\mathbf{b}$ axes, as illustrated in Fig.~\ref{fig:DM_Frustration}~(b) and (c). Sign-alternating Dzyaloshinskii vectors are typically responsible for the phenomenon of {\em weak ferromagnetism} (WFM) \cite{Dzyaloshinsky1958, Moriya1960}. Not every antiferromagnetic structure will develop WFM. In our model (see Fig.~\ref{fig:DM_Frustration}~(b)) the \CAFb phase {\em does not}, because the direction of $\mathbf{D}$-induced  canting (the sign of $\mathbf{D}\cdot(\mathbf{S}_1\times \mathbf{S}_2)$) alternates from bond to bond, resulting in zero net magnetization. On the other hand, for the \CAFa structure the $\mathbf{D}$ vectors are in sync with the spin cross-products (see Fig.~\ref{fig:DM_Frustration}~(c)), generating a net WFM magnetization. Thanks to extra Zeeman energy, a WFM state is obviously favored by an applied field, which eventually results in a \CAFb$\rightarrow$\CAFa transition. To make this quantitative, lets compare the classical exchange energy of the two phases, expressing it as a function of $\theta$, the $(a,c)$-plane angle between sublattice spins and field direction. In the limit $D\ll J$, we can assume that $\theta$ is defined by the balance of Zeeman and Heisenberg-exchange energies alone, and is the same for all spins. The classical energy difference per spin for the two phases is:
		\begin{equation}
			E_\text{a}-E_\text{b}=S^2[1-\cos(2\theta)]\,\left[J_{1,2}-J_{1,1}\right]+DS^2\sin(2\theta).
		\end{equation}
		The first term is quadratic with $\theta$ and will always lose out to the second (linear) term for small enough $\theta$, i.e., close enough to saturation. For {\em arbitrarily small} $\mb{D}$, close enough to saturation,  the system will switch from \CAFb to \CAFa. To see if this scenario makes physical sense for \SrZn, we can estimate the required magnitude of $\mathbf{D}$. The transition is observed at a magnetization $m\sim 0.92 \,m_\text{sat}$, which corresponds to $2\theta\sim 0.26\pi$. Then the classical energies of the two states are equal for $D\sim 0.4 (J_{1,1}-J_{1,2})\sim 0.04$~meV. That is an entirely reasonable magnitude of off-diagonal exchange for $J_{2,1}\sim 0.9$~meV and completely in-line with our previous estimate $D\sim 0.05$~meV based on the value of the spin flop field for $\mathbf{H}\|\mathbf{a}$ \cite{Landolt2021}. This said, we once again emphasize that this can {\em not} be the whole story, since in the presence of an external field a ``ferromagnet'' is nothing else but  the paramagnetic  phase. The toy model entirely fails to explain the $H_\text{sat}$ transition.
		
		Any new  $(h,k,0)$ Bragg reflections with non-integer indexes, if present, would have likely been detected in the TOF data. Since they are absent, we conclude that the presaturation state may be a different  $\mathbf{Q}=0$ structure (with weak magnetic peaks hiding underneath nuclear ones). Alternatively, it may have  substantial propagation vector component along the $\mathbf{c}^\ast$ direction. For example, it could feature $(h,k,[\text{odd}])$-type Bragg peaks, corresponding to an anti-parallel alignment of spins in adjacent \VV-layers. Unfortunately, testing this hypothesis is not technically feasible: in any realistic neutron experiments applying a 14~T field along the $c$ axis will constrain scattering to the $(h,k,0)$ plane, due to the use of split-coil cryomagnets.

		\subsection{Quantum corrections to spin wave theory}
		
		Regardless of the nature of the presaturation state, our data nicely highlight the importance of quantum corrections to spin wave theory for \SrZn. This is already apparent in the highly non-linear magnetization curve, as compared to a straight line in the classical  Heisenberg model  (Fig.~\ref{fig:Magnetization}, dashed line). The non-linearity can be to some extent accounted for already by the first order $1/S$ correction.  The red solid line in Fig.~\ref{fig:D23_FieldScan} is such a calculation for a $J_1$-$J_2$ square lattice model \cite{Thalmeier2008}. Here we used $J_1=J_{1,1}$ and $2J_2=J_{2,1}+J_{2,2}$ to ensure the correct value of the saturation field. Some discrepancies with experiment persist in low fields, where higher-order $1/S$ terms are known to become considerably more important \cite{Thalmeier2008}.
		
		Quantum corrections to SWT are even more obvious in the differences of exchange constants obtained in neutron experiments at zero field and above saturation. To within the resolution of all our inelastic measurements the anisotropy responsible for the spin flop and presaturation transitions is too small to be relevant. We can therefore discuss the measured spectrum in the context of the Heisenberg model. Since SWT becomes {\em exact} for the quantum Heisenberg model in the fully polarized phase, it is actually the high-field values that represent correct parameters of the microscopic Hamiltonian. In contrast, the zero-field values are ``renormalized'' effective parameters. Note that while most  exchange parameters are renormalized downwards, the strongest one is renormalized up, by as much as 40~\%. It is worthy to note that this renormalization can be estimated solely with the zero field exchange couplings by comparing the classically expected saturation field with actual one. In case of \SrZn the zero field exchange constants overestimate the saturation field by 40\% \cite{Landolt2021}. A similar situation can be found in \PbV where the expected saturation field is 28.6~T, again about 40\% larger than the observed saturation field of 20.7~T \cite{Landolt2020} \footnote{The statement to the contrary in \cite{Landolt2021} is erroneous and due to a calculation mistake}. This is approaching the extreme renormalization in the one-dimensional Heisenberg spin chain, where the sharp bound of the spinon continuum, aka the De Cloizeaux-Pearson ``spin wave'' \cite{Cloizeaux1962}, has a bandwidth that is $\pi/2$ times larger than the actual exchange constant \cite{Mourigal2013a}. 
		
		Deviations from SWT are visible in the excitation spectrum not only at zero field, but even just below saturation. This is  already noted in the context of intensity distribution between the two spin wave branches. In the vicinity of $(0,1,0)$ the branch which disappears at saturation corresponds to ``optical'' oscillations of the ordered staggered magnetization. The corresponding structure factor is directly linked to the magnitude of the latter and thus to the canting angle $\theta$. That, in turn, is defined by the deviation of magnetization from saturation. SWT's failure to correctly reproduce the intensity of this mode is therefore a direct consequence of SWT incorrectly predicting magnetization. To illustrate this, in Fig.~\ref{fig:HYSPEC_QCuts}, in solid lines we show an SWT simulation at  fictitious values of magnetic field chosen such that, if plugged into SWT, would predict the correct magnetization as actually measured experimentally. From the experimental magnetization data in \ref{fig:Magnetization}, we conclude that for $\mu_0H=13.8$~T and $\mu_0H=13.4$~T these fictitious fields are $13.3$~T and $12.3$~T, respectively. The corresponding simulations in Fig.~\ref{fig:HYSPEC_QCuts} are obviously off in excitation energy, but reproduce the experimentally observed intensity balance between the two branches rather well.

		\section{Conclusion}
		Due to its frustrated and quasi-two-dimensional nature, spin waves and the magnetization process in \SrZn are subject to very strong quantum corrections. At the same time, the disappearance of the columnar-antiferromagnetic structure in a discontinuous presaturation transition is probably of a purely classical origin and caused by Dzyaloshinskii-Moriya interactions.  However, the nature of the order parameter in the presaturation state remains undetermined.

		\acknowledgments
		This work is partially supported by the Swiss National Science Foundation under Division II. This work was additionally supported by the Swiss State Secretariat for Education, Research and Innovation (SERI) through a CRG-grant. The neutron scattering data collected on IN12 for the present work are available at \href{https://doi.ill.fr/10.5291/ILL-DATA.CRG-2717}{https://doi.ill.fr/10.5291/ILL-DATA.CRG-2717}. A portion of this research used resources at the Spallation Neutron Source, a DOE Office of Science User Facility operated by the Oak Ridge National Laboratory. Andrey Zheludev thanks Dr. Mike Zhitomirsky (ILL) for enlightening discussions.

		\bibliography{SrZn-bib}
		
	\end{document}